\documentclass[aps,prl,reprint]{revtex4-1}
\usepackage{amsmath,amssymb}
\usepackage{graphicx}
\usepackage{color}
\usepackage{bm}
\usepackage[colorlinks,linkcolor=blue]{hyperref}

\newcommand{\dfdx}[2]{\left(\frac{\partial #1}{\partial #2}\right)}
\begin{document}
\title{Pressure Gradients Fail to Predict Diffusio-Osmosis}\label{phoresis}

\author{Yawei Liu} 
\affiliation{Beijing Advanced Innovation Center for Soft Matter Science and Engineering, Beijing University of Chemical Technology, Beijing 100029, China}
\author{Raman Ganti}
\affiliation{Department of Chemistry, University of Cambridge, Lensfield Road, Cambridge CB2 1EW, UK}
\author{Daan Frenkel}
\email[Corresponding author: ]{df246@cam.ac.uk}
\affiliation{Department of Chemistry, University of Cambridge, Lensfield Road, Cambridge CB2 1EW, UK}

\date{\today}

\begin{abstract}
We present numerical simulations of diffusio-osmotic flow, i.e. the fluid flow generated by a concentration gradient along a solid-fluid interface.  In our study, we compare a number of distinct approaches that have been proposed for computing such flows and compare them with a reference calculation based on direct, non-equilibrium Molecular Dynamics  simulations. As alternatives, we consider schemes that compute diffusio-osmotic flow from the gradient of the chemical potentials of the constituent species and from the gradient of the component of the stress tensor parallel to the interface. We find that the approach based on treating chemical potential gradients as external forces acting on various species agrees with the direct simulations, thereby supporting the approach of  Marbach et al. (J Chem Phys 146, 194701 (2017)).  In contrast, an approach based on computing  the gradients of the microscopic pressure tensor does not reproduce the direct non-equilibrium results. 

\end{abstract}

\maketitle

\section{Introduction}
Flow in macroscopic channels is driven by pressure gradients or body forces, such as gravity. In contrast, flow in nano-structured materials (e.g. nano-channels), is usually dominated by phoretic effects, where transport is caused by thermodynamic gradients acting near interfaces. The key point to note is that the gradients responsible for phoretic flow (e.g. electrical fields, concentration gradients or thermal gradients), cannot cause bulk flow: they only act in narrow interfacial layers where the fluid experiences surface-specific interactions. In view of the increasing importance of  micro- and nano-fluidic devices and self-propelling particles (see, e.g.~\onlinecite{Stone2004, Squires2005, Bocquet2010a, Bocquet2014, Golestanian2005, Paxton2006, Guix2014}) it becomes important to be able to predict the strength of phoretic flow phenomena based on knowledge of the microscopic interactions between the atoms or molecules in the system. In the present paper, we focus on the numerical prediction of  diffusio-osmosis, where a chemical potential gradient along a solid surface drives the  flow~\cite{Anderson1982,Anderson1989,Brady2011a}. 
Traditionally, such flows have been described using a continuum picture, where the material properties were characterised by (local) thermodynamic quantities and the flow was computed, assuming that the (Navier-)Stokes equation holds near the surface. Such a macroscopic perspective is appropriate in systems where the fluid-interface interaction acts over a range that is much larger than a typical molecular size, as is the case for electro-osmosis of dilute electrolytes. However, as most inter-molecular forces have a range comparable to a molecular diameter, the continuum picture is not expected to hold for most diffusio-osmotic phenomena.

In what follows, we focus on a situation where the surface-fluid interaction is short-ranged. Specifically, we consider MD simulations of diffusio-osmosis of a neutral solvent containing a neutral solute, near a  crystalline  solid surface that is, on average, flat. 

Thus, to determine the flow velocity, it is critical to accurately calculate the 
surface force induced by the concentration gradient. The flow can then be obtained from MD simulations by applying the force 
to fluid particles. 
In order to validate our method, we developed an ingenious non-equilibrium simulation technique to directly compute the flow.

\section{ Thermodynamic gradients}
We first consider an atomically flat wall at $z(x,y)\le 0$, in contact with a fluid mixture in the region $z>0$. The fluid contains a majority component ($A$: `solvent') and a minority component ($B$: `solute'). The fluid, as a whole, is maintained at constant bulk pressure and constant temperature. When a concentration gradient in $B$ (and, via the Gibbs-Duhem relation, also in $A$) is imposed in the fluid along $x$, flow occurs due to a pressure gradient  at the interface. As the concentration gradients in $A$ and $B$ are not independent, we will focus on the phoretic effect of the concentration gradient of $B$. It should be borne in mind that this effect, includes the effect of the concentration gradient in $A$. The most intuitive (but, as we will show, incorrect) method to obtain the  force acting on fluid particles near an interface is to calculate the force  due to the pressure gradient acting  on a small volume element, and then obtain the force per particle by dividing the force per volume by the local number density. The transverse component of the pressure tensor at $z$ ($p^{xx}(z)$) depends on $x$ only through its dependence on the spatial variation of the bulk concentration ($\rho_B$) or, equivalently, of the chemical potential ($\mu_{B}$) of the species subject to a concentration gradient:
\begin{equation}\label{eq:eq01}
f^{V}(z) = -\frac{\partial p^{xx}(z)}{\partial \rho_B}  \frac{\partial \rho_B}{\partial x}.
\end{equation}
In the case of a sufficiently small concentration gradient, we can assume local thermodynamic equilibrium (LTE) -- deviations from LTE are expected to be of higher than linear order in the concentration gradient. Thus, rather than computing the local pressure tensor in a non-equilibrium system, we can compute it in equilibrium as a function of  concentration. The pressure gradient is then computed using
\begin{multline}\label{eq:eq02}
f^{V}(z) = -\frac{\partial p^{xx}(z)}{\partial \rho_B}  \frac{\partial \rho_B}{\partial x} \\
\approx  -\frac {p^{xx}_{\rho_B+\Delta \rho_B}(z)-p^{xx}_{\rho_B-\Delta \rho_B}(z)} {2\Delta \rho_B} \cdot \nabla \rho_B,
\end{multline}
where $\Delta \rho_B$ is a small change in the bulk concentration of species $B$. Note that we will always assume that the bulk pressure is constant.

The local pressure tensor at position $\bf r$ is defined as the ensemble average of the negative of the stress tensor at $\bf r$~\cite{Irving1950,Schofield1982}. It contains two terms: a kinetic term, arising from the change in momentum due to particles crossing the boundaries of an elemental  volume at $\bf r$, and a configurational term, related to the change in momentum due to  intermolecular interactions between the particles. There is, however, a problem:  for inhomogeneous systems (e.g. fluid near a solid wall), the configurational component of the pressure tensor cannot be defined uniquely. For the computation of surface tension, this ambiguity has no effect~\cite{Schofield1982}, but for the computation of phoretic flow, the problem does not go away.

In the present work, we explore the predicted phoretic flow for two different definitions of the local pressure tensor~\cite{Kirkwood1949,Harasima1953} and use these definitions to compute  the transverse pressure as a function of $z$ in a series of slabs parallel to the wall. We employed both the Irving-Kirkwood definition, in which an intermolecular force contributes to the local pressure in every slab between two molecules~\cite{Irving1950}, and the Virial definition in which an intermolecular force contributes to the local pressure in the slab(s) where the two molecules are located~\cite{Weng2000,Cormier2001}. 


Alternatively, we can use local thermodynamics to compute the force driving flow. Consider an $n$-component fluid mixture at constant temperature, $T$. The Gibbs-Duhem relation can be written as $Vdp = \sum_{i=1}^{n} N_i d \mu_i$ where  $N_i$ is the number of particles of species $i$ in volume $V$, $p$ the pressure and $\mu_i$ the chemical potential of species $i$. Let us denote the number density of species $i$ in the mixture by $\rho_i$. Then, $dp = \sum_{i=1}^{n} \rho_i d \mu_i$. A concentration gradient of  component $i$ along $x$ will lead to a chemical potential gradient $\partial \mu_i / \partial x$. As the pressure remains constant in the bulk, Gibbs-Duhem relation reduces to $0= \sum_{i=1}^n \rho_i^\text{bulk}(x) \left( \partial \mu_i / \partial x \right)$. At a position $z$ near the interface, a pressure gradient remains, giving a force per unit volume
\begin{multline}\label{eq:eq03}
f^V(z)=\left( -\frac {\partial p(z,x)} {\partial x} \right) \\
= \sum_{i=1}^n \left( \rho_i(z,x)-\rho_i^\text{bulk}(x) \right) \left( -\frac {\partial \mu_i} {\partial x} \right).
\end{multline}
We can interpret $\left( -{\partial \mu_i} / {\partial x} \right)$ as the force  per-particle acting on the particles of species $i$. This expression is convenient, because the imposed chemical potential gradients are constant throughout the system. In the bulk, the composition is such that the forces balance (the bulk pressure equilibrates rapidly). Upon approaching the wall, the concentration of different components changes, leading to non-zero net forces. In other words,  particles of a given species experience the same force regardless of their distance from the interface. The force acting on species $i$ is then
\begin{equation}\label{eq:eq04}
f_i = \left( -\frac {\partial \mu_i^\text{bulk}} {\partial x} \right) =\left( -\frac {\partial \mu_i^\text{bulk}} {\partial \rho_i} \right)_{P} \cdot \nabla \rho_i.
\end{equation}
We now have two approaches (Eq.~\ref{eq:eq02} and \ref{eq:eq03}) for computing the force driving diffusio-osmotic flow. Eq.~\ref{eq:eq02} is a mechanical expression while Eq.~\ref{eq:eq03} is thermodynamic.  Of course, in steady state, these forces are exactly balanced by the force due to the gradient in the shear stress in the moving fluid. To test which, if either, of these microscopic expressions is correct, we performed MD simulations on the simple model system mentioned above. 

\section{Molecular dynamics simulations}
\begin{figure}[tb]
	\begin{center}
		\includegraphics[width=0.45\textwidth]{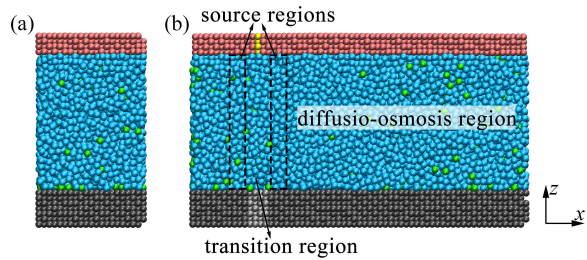}
		\caption{(a) Simulation box used to compute the force and flow profiles via the forces obtained from Eq.~\ref{eq:eq02} and Eq.~\ref{eq:eq04}. (b) Simulation box used in the direct non-equilibrium MD simulations with explicitly imposed concentration gradients. The blue particles represent the solvent ($A$), the green particles represent the solute ($B$), the red and yellow particles represent the solid particles in the top wall, and the black and silver particles represent the solid particles in the bottom wall.}
		\label{fig:fig01}
	\end{center}
\end{figure}

\begin{figure*}[tb]
	\begin{center}
        \includegraphics[width=0.8\textwidth]{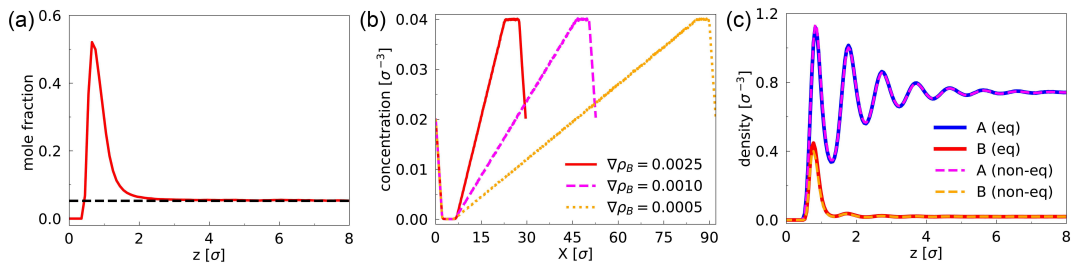}
		\caption{(a) The solute mole fraction profile along $z$ from an equilibrium simulation at $\rho_B=0.04$. (b) The bulk concentration profiles in the non-equilibrium simulations.  (c) The average density profiles in the diffusio-osmosis region in the direct non-equilibrium simulation at $\nabla \rho_B =0.0025$, and the density profiles from the equilibrium simulation at $\rho_B=0.02$.}
		\label{fig:fig02}
	\end{center}
\end{figure*}

We performed MD simulations of a fluid mixture composed of solvent ($A$) + solute ($B$) particles, confined between two parallel solid walls (Fig.~\ref{fig:fig01}). All particles have the same molecular diameter $\sigma$. In what follows, we use $\sigma$ as our unit of length. Interactions between these particles are given by a Lennard-Jones potential truncated and shifted at $4\sigma$, $U_{\alpha\beta}(r) = 4 \epsilon_{\alpha\beta} \left[ (\sigma / r)^{12} - (\sigma / r)^6 \right]$ ($\alpha, \beta \in \left\{ A,B,\text{top},\text{bottom} \right\}$), with $\epsilon$ interaction energy. To narrow our exploration, we focused on the ideal solution composed of identical solvent and solute particles in the bulk, but different values of $\epsilon$ with the bottom wall. We chose 
$\epsilon_{AA} = \epsilon_{BB} = \epsilon_{AB} = \epsilon_{A,top}=\epsilon_{B,top} \equiv  1.0  \epsilon$ and $\epsilon_{B,bottom}= 2\epsilon_{A,bottom}=  1.1  \epsilon$. Thus, the pressure difference only appears in the fluid near the bottom wall when a concentration gradient is imposed. In the remainder of this paper, we use $\epsilon$ as our unit of energy and the mass $m$ of the fluid particles as our unit of mass.

All simulations were carried out using the LAMMPS package~\cite{Plimpton1995}  in an isothermal, isobaric ($NP^{zz}T$) ensemble. The top wall was used as a barostat and was otherwise free to move in the horizontal and vertical direction. A constant external force along $z$ was exerted on the top wall to maintain the bulk pressure at a value $p_{ex}=0.012$. Periodic boundary conditions were imposed in the $x$ and $y$ directions. The solid walls were composed of  solid particles placed on a FCC lattice with lattice spacing of $1.64$ and each particle was connected to its nearest neighbours via stiff harmonic bonds with spring constants  2500$\epsilon/\sigma^2$ equilibrium length  $1.64/\sqrt{2}\sigma$.
During the simulations, the layer of solid particles in the bottom wall furthest removed from the interface was rigidly anchored.  The velocity-Verlet algorithm with a time step of $0.001$ was used to integrate the equations of motion, and a Nos\'e-Hoover thermostat with a time constant of $0.1$ was used to maintain the temperature at $T=0.846$.  All simulations were run for $2\times10^8 - 4\times10^8$ steps to obtain sufficient statistics.

The calculation of $\partial p^{xx}(z)/\partial \rho_B$ required several equilibrium simulations at different uniform bulk concentrations. These simulations could be carried out in a relatively small simulation box as shown in Figure~\ref{fig:fig01}(a). The box dimensions were  $L_x=16.44$, $L_y=9.86$, and $\langle L_z \rangle=29.7$ ($L_z$ fluctuates, and depends on the solute concentration). The system contained  $2640$ fluid particles. To compute the composition-dependence of $p^{xx}(z)$, we performed simulations where we varied the concentration of the solute $B$, while keeping the total number of particles fixed.
From the numerical estimate of $\partial p^{xx}(z)/\partial \rho_B$, we computed the corresponding force using Eq.~\ref{eq:eq02} with $\rho_B = 0.02$, $\Delta \rho_B=0.01$, and various $\nabla \rho_B$. We verified that our estimate of the pressure gradient did not depend on the size of $\Delta \rho_B$.
Subsequently, we converted the force per unit volume to a force per particle, by dividing by the total number density profile $\rho(z)$. These per-particle forces were then applied in a non-equilibrium simulation of the fluid with solute density  $\rho_B$ to determine the flow profile for a given $\nabla \rho_B$ [Fig.~\ref{fig:fig01}(a)].
Similarly, starting from Eq.~\ref{eq:eq04}, we can compute the forces that would result from the gradient of the chemical potentials. These per-particle forces were applied to the solute and solvent particles.

To validate our approach, we performed direct non-equilibrium simulations where a concentration gradient was explicitly imposed.  Figure~\ref{fig:fig01}(b) shows the simulation box in which the fluid mixture has a constant bulk concentration gradient along $x$. 
Non-equilibrium simulations were carried out to measure the flow profile at different values of $\nabla \rho_B$. In this case, we employed boxes that contain several regions: two source regions with the width of $4\sigma$, a diffusio-osmosis region with various widths, and a transition region of width $\sim8\sigma$ between the two source regions. During the simulations, every $500$ steps, the identities of the fluid particles in these source regions were reset to maintain constant concentrations and a steady gradient along $x$. In the low concentration source region, $\rho_B = 0$ so that all fluid particles were reset to the solvent type. In the high concentration source region, $\rho_B = 0.04$. As the concentration varies only close to the wall, the fraction of selected particles reset to the solute type in each slab parallel to the wall is equal to the local solute mole fraction calculated via an equilibrium simulation at $\rho_B=0.04$ [Fig.\ref{fig:fig02}(a)]. In the transition region, we set $\epsilon_{A,bottom}=\epsilon_{B,bottom}=0.55$ to prevent diffusio-osmosis in this region. A steady flow can be achieved with an ingenious design: During all simulations, the bottom wall is fixed by freezing particles in the last layer of the bottom wall. As the top wall is free, it moves at the same velocity as the bulk fluid. We tracked the position of the top wall (i.e. the position of yellow particles in the top wall) and redefined the position of source regions each time the identities of fluid particles are reset. The concentration gradient along $x$ depends on the box size. In this work, the box size is $L_x=36.2$, $52.6$, and $92.1$, $L_y=9.86$, and $\langle L_z \rangle=29.7$ for three independent simulations. The box dimensions correspond to $5808$, $8448$, and $14784$ fluid particles, respectively.

Figure~\ref{fig:fig02}(b) shows the bulk concentration gradient along $x$ for the non-equilibrium simulations. The figure shows that the concentration profile is linear between the limiting values imposed in the source regions. The concentration gradients in the three independent simulations were  $\nabla \rho_B =0.0025$, $0.0010$ and $0.0005$, respectively. Figure~\ref{fig:fig02}(c) shows the density profiles for each component along $z$ from the simulation of $\nabla \rho_B =0.0025$. As the average bulk concentration is $0.02$, the density profiles from the equilibrium simulation at $\rho_B=0.02$ were also plotted for comparison. The results show good agreement for the local densities from the two simulations, indicating that in the non-equilibrium simulation, fluid states are still close to equilibrium, validating the  assumptions underlying our calculation of the surface force via Eqs.~\ref{eq:eq01} and \ref{eq:eq03}.

\section{Results and discussion}

\begin{figure*}[tb]
	\begin{center}
		\includegraphics[width=0.8\textwidth]{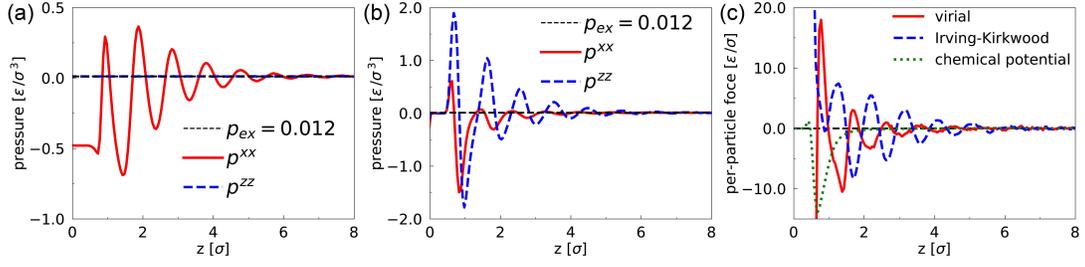}
		\caption{(a) The Irving-Kirkwood pressure profiles at $\rho_B=0.01$. (b) The virial pressure profiles at $\rho_B=0.01$.  (c) The average per-particle force profiles at $\rho_B=0.02$ and $\nabla \rho_B=1.0$.} 
		\label{fig:fig03}
	\end{center}
\end{figure*}

\begin{figure}[tb]
	\begin{center}
		\includegraphics[width=0.45\textwidth]{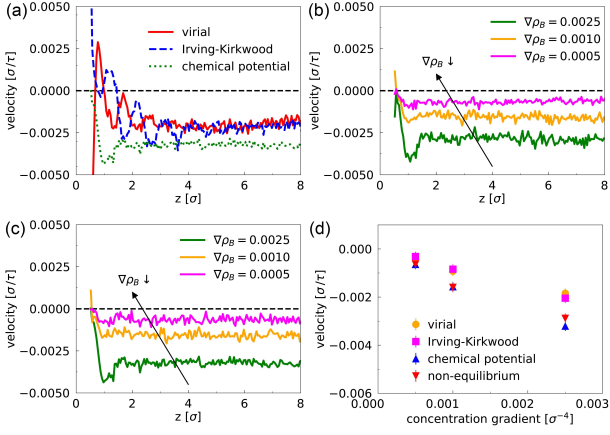}
		\caption{(a) The flow velocity profiles at $\rho_B=0.02$ and $\nabla \rho_B=0.0025$ from the simulations with applying the surface forces computed from different methods. (c) The flow velocity profiles measured from the direct non-equilibrium simulations. (d) The slip velocity at different concentration gradients from different methods.}
		\label{fig:fig04}
	\end{center}
\end{figure}

\begin{figure*}[tb]
	\begin{center}
		\includegraphics[width=0.8\textwidth]{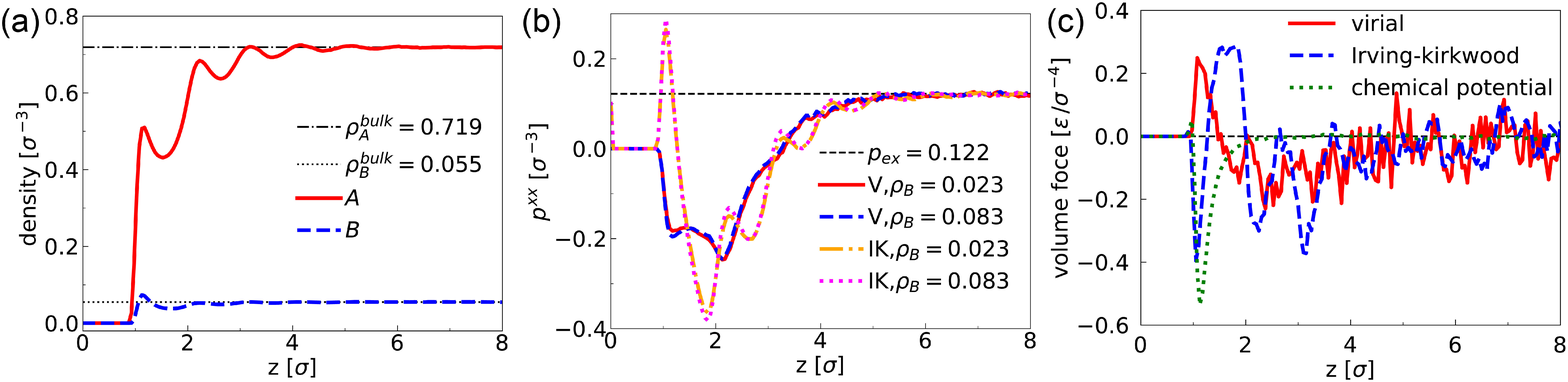}
		\caption{(a) Density profiles for species $A$ and $B$ at $\rho_B=0.055$. (b) Virial (V) and Irving-Kirkwood (IK) pressure profiles at $\rho_B=0.023$ and $\rho_B=0.083$. (c) The average volume force profiles calculated for $\nabla \rho_B=1.0$.} 
		\label{fig:specular_wall}
	\end{center}
\end{figure*}

In order to calculate the surface force at $\rho_B = 0.02$ via Eq.~\ref{eq:eq02}, we computed the pressure-tensor profile at $\rho_B = 0.01$ and  $\rho_B = 0.03$. Figures~\ref{fig:fig03}(a) and (b) show the pressure profiles along $z$ near the wall at $\rho_B=0.01$ using the Irving-Kirkwood and virial definitions. In the bulk, where fluid is homogeneous and far away from the wall, all definitions lead to the same value since $p^{zz}=p^{xx}=p_{ex}=0.012$. Upon approaching the wall, $p^{zz}$ from the Irving-Kirkwood definition is, as expected, equivalent to the bulk pressure, reflecting mechanical equilibrium along $z$ [Fig.~\ref{fig:fig03} (a)], while the virial expression for $p^{zz}$ is not constant  [Fig.~\ref{fig:fig03} (b)]. For $p^{xx}$, the different expressions for pressure result in  different, oscillating profiles near the wall.

The chemical potential for component $i$ is given by $\mu_i = \mu_i^0 + k_BT \ln \rho_i^{bulk} + \mu_i^{exc}$, with $k_B$ the Boltzmann constant. $\mu_i^0$ denotes a (constant) reference value and  $\mu_i^{exc}$ denotes the excess chemical potential due to intermolecular interactions. Because the bulk solutions  are ideal, $\mu_i^{exc}$ does not depend on the concentration of $B$. Thus, at $\rho_B = 0.02$, with $\rho_A^{bulk}=0.74$ and $\rho_B^{bulk}=0.02$ [Fig.~\ref{fig:fig02}(b)], if $\nabla \rho_B=1.0$, we obtain $f_A=1.14$ and $f_B=-41.75$.

We can now compare the force profile from the pressure gradients with those from the chemical potential gradients. Figure~\ref{fig:fig03}(c) shows the average per-particle force acting on the fluid particles at $\rho_B=0.02$ and $\nabla \rho_B=1.0$ for all of the methods. For the chemical potential method, the average per-particle force is $f^{ave}(z) = [\rho_A(z) f_A + \rho_B(z) f_B] / \rho(z)$. As shown in the figure, the two expressions for the surface force (Eq.~\ref{eq:eq01} and \ref{eq:eq03}) produce significantly different results near the interface. Unsurprisingly, the forces calculated from the chemical potential gradients (Eq.~\ref{eq:eq03}) are concentrated where there is an excess of solute [Fig.~\ref{fig:fig02}(b)]. However, the forces calculated via the local pressure tensors i.e. the Irving-Kirkwood and virial definitions (Eq.~\ref{eq:eq01}), extend over larger distances and  fluctuate strongly.  

The fact that pressure tensor and chemical potential routes yield different force profiles implies that they would predict different flow profiles. At most, one of these can be correct. To test whether the computed  flow profiles are correct, we applied the force profiles that we computed to the fluid mixture at $\rho_B=0.02$ and measured the flow velocity as a function of $z$ for fixed $\nabla \rho_B$. Figure~\ref{fig:fig04}(a) shows the predicted velocity profiles at $\nabla \rho_B=0.0025$. We see that the velocity profiles from different methods are significantly different. To validate the chemical potential and pressure-gradient calculations, the velocity profile at the same $\nabla \rho_B$ was also computed directly in a non-equilibrium simulation [Fig.~\ref{fig:fig01}(b)]. The result is shown in Fig.~\ref{fig:fig04}(b). We see that the velocity profile that follows from the direct simulation differs markedly from the one obtained from the pressure gradients. However, it agrees quite well with the predictions based on the chemical-potential gradients. The latter agreement was also observed for two other concentration gradients ($\nabla \rho_B=0.0010$ and $0.0005$) [Fig.~\ref{fig:fig04}(b) and (c)]. 

We also compared the  slip velocity (i.e. the velocity of the bulk fluid with respect to the wall, $v_s$) obtained from different methods. The results are shown in Fig.~\ref{fig:fig04}(d). We see that prediction of $v_s$ based on the chemical-potential gradients is in excellent agreement with those obtained from direct simulations. However, $v_s$ predicted based on the pressure gradients is significantly different.

The failure of mechanical expressions near the interface is consistent with our recent calculations on the solutal Marangoni effect~\cite{liu2017microscopic} and thermo-osmosis~\cite{ganti2017hamiltonian}. Therefore, it is not entirely surprising that the method also fails here. Yet, where the present result differs from our earlier studies 
is that the pressure gradient method also predicts an incorrect value of the bulk velocity [Fig.~\ref{fig:fig04}(d)]. More interestingly, the flow velocity profile computed with the (presumably correct) chemical-potential-gradient method shows an overshoot near the wall (the effect is clearest for larger solute concentrations). This observation is interesting because if the flow profile could be computed from the force profile using the Stokes equation (i.e. assuming a position independent viscosity) then an overshoot is not possible if  the force always has the same sign (as it does -- see Fig.~\ref{fig:fig03}(c), green dotted curve).  

To understand the failure of the pressure-gradient approach, it is useful to revisit the thermodynamic description of diffusio-osmotic transport. On a macroscopic level, it is the gradient in the surface free-energy density $\gamma$ (which for fluid-fluid interfaces is equal to the surface tension) that determines the flow: in the case of fluid interfaces, this is the well-known Marangoni effect~\cite{liu2017microscopic}. 

If we consider the variation of $\gamma$ with the chemical potential of the species in a binary mixture, we can write
\begin{eqnarray}
\label{eq:gibbs_adsorption_st_grad}
\dfdx{\gamma}{x}_{P,T}&=&\dfdx{\gamma}{\mu_B}\dfdx{\mu_B}{x}+\dfdx{\gamma}{\mu_A}\dfdx{\mu_A}{x}\nonumber\\
&=&-\left[ \Gamma_B \dfdx{\mu_B}{x}+\Gamma_A \dfdx{\mu_A}{x}\right]
\end{eqnarray}
where $\Gamma_{i}$ ($i$= $A$ or $B$) is the Gibbs adsorption of species $i$ at the interface.
It is clear that this thermodynamic expression immediately yields a relation between the driving force of the Marangoni flow and the chemical potential gradients, which is in agreement with our observation that the microscopic expression [Eq.~\eqref{eq:eq03}] couples the chemical potential gradient to the local excess density of the corresponding species.

For a liquid-liquid (or a liquid-flat wall) interface, we can use the Kirkwood and Buff expression to relate the surface tension to the pressure tensor
\begin{equation}
\label{eq:conventional_surface_tension}
\gamma = \int_{-\infty}^{\infty} p - p^{xx}(z) \: \mathrm{d}z
\end{equation}
where $p$ is the hydrostatic pressure. Differentiating Eq.~\eqref{eq:conventional_surface_tension} at constant pressure and temperature gives
\begin{equation}\label{eq:mechanical_st_grad}
\dfdx{\gamma}{x}_{P,T} = - \int_{-\infty}^{\infty} \dfdx{p^{xx}(z)}{x} \: \mathrm{d}z.
\end{equation}
And hence, the driving force for flow is, in that case, related to the gradient of the pressure tensor. Yet, when the solid surface is not flat, but atomistically structured, Eq.~\eqref{eq:conventional_surface_tension} no longer holds. The latter integral would yield the surface {\em stress}. The gradient of the surface stress is not the driving force for particle transport, and hence for structured walls, we should not expect to obtain the driving force for diffusio-osmotic flows from stress gradients. 

The previous arguments can be numerically tested by multiplying the per-particle force profiles shown in Fig.~\ref{fig:fig03}(c) by $\rho(z)$ and integrating from the surface into the bulk. The surface tension gradient predicted by Eq.~\eqref{eq:mechanical_st_grad} using the virial and Irving-Kirkwood pressure expressions is 1.07 as opposed to -5.81 predicted by Eq.~ \eqref{eq:gibbs_adsorption_st_grad}. These numerical results confirm that for a structured surface, the stress gradient is not the driving force.

For an ideally flat wall, however, we would expect Eq.~\eqref{eq:mechanical_st_grad} to predict accurately the driving force. We investigate the latter case by applying our methods to fluid interacting with a surface via specular boundary conditions. Fluid-fluid interactions remain unchanged, but fluid-wall interactions are governed by
\begin{align}
U_{fw}(z) = 4 \epsilon_{fw} \left[ (\sigma / z)^{12} - (\sigma / z)^6 \right]
\end{align}
where $\epsilon_{B,w} = 2\epsilon_{A,w} = 1.1\epsilon$. In this system, the wall is simply a flat surface located at $z=0$. Therefore, fluid atoms at a position $z$ near the surface will experience the same wall force for all $x$ and $y$. Changing the wall potential required increasing the external pressure to $p_{ex} = 0.122$ to maintain a stable system.

In order to achieve sufficient signal via the pressure gradient approach, $\Delta \rho_B$ was increased to $0.03$ in Eq.~\eqref{eq:eq02}. Fig~\ref{fig:specular_wall}(a) and (b) show density and pressure profiles for fluid interacting with the specular wall. Comparing with Fig~\ref{fig:fig03}(a) and (b), it is clear that removing transverse force contributions from the wall significantly changes the behavior of fluid near the interface.

The surface tension gradient can be computed by integrating the per-volume force profiles shown in Fig~\ref{fig:specular_wall}(c) from $z=0$ to $z=5$, beyond which none of the methods predict any effect. The results were $-0.19 \pm 0.01$ using Eq.~ \eqref{eq:gibbs_adsorption_st_grad}, $-0.2 \pm 0.1$ and $-0.26 \pm 0.1$ using the virial and Irving-Kirkwood expressions in Eq.~\eqref{eq:mechanical_st_grad}.  Clearly, the forces that follow from the stress gradients are subject to considerable statistical noise. As a consequence, the only thing we can say is that there the differences between the integrated forces are not statistically significant. This observation is in agreement with our earlier results for the Marangoni flow near a (flat) liquid-liquid interface~ \cite{liu2017microscopic}. 

However, as the predicted force profiles are very different for the virial and Irving-Kirkwood expressions, and as both are different from the force-profile obtained from the chemical potential gradient,  we do expect that the flow profiles that would result from solving the Stokes equation with non-slip boundary conditions would be different. In that case, only the flow profile resulting from chemical potential gradients is to be trusted.  In view of the poor statistical accuracy of the computed stress gradients, computing the flow profiles explicitly would have been meaningless.

We also note that using the Stokes equation would not be justified near a wall, because  the viscosity is not expected to be constant, due to layering of the fluid near the wall.  Computationally, this is not a problem, because we can compute the flow velocity directly, as in Fig.~\ref{fig:fig04}.

The numerical results presented here show that if a solid surface exerts non-zero transverse stress on the fluid, standard pressure expressions fail to predict the surface tension gradient via Eq.~\eqref{eq:mechanical_st_grad}. Still, we can proceed further. Following Schofield and Henderson's analysis of the microscopic pressure tensor, the difference between standard and non-standard expressions lies in choosing the contour along which the intermolecular force is integrated~\cite{schofield1982statistical}. Irving and Kirkwood consider a straight line connecting atom pairs, though this choice is arbitrary. Yet, even if we formulated the pressure using a different contour, the transverse force contributions from the solid surface would result in incorrect predictions of the surface tension gradient. Therefore, in the case of a structured solid surface, any pressure expression will microscopically and macroscopically fail to predict the surface force that drives diffusio-osmotic flow.



\begin{acknowledgments}
We gratefully acknowledge discussions with Lyd\'{e}ric Bocquet, Mike Cates, Patrick Warren, Ignacio Pagonabarraga and Benjamin Rotenberg. YL would like to acknowledge the hospitality of the Chemistry Department of the University of Cambridge. RG gratefully acknowledges a PhD Grant from the Sackler Fund.  DF  acknowledges support by the European Union through the European Training Network NANOTRANS Grant 674979. 
\end{acknowledgments}

\bibliography{text}

\end{document}